\input psfig
\documentstyle[aas2pp4,flushrt]{article}
\begin{document}
\title{Secondary CMB anisotropies in a universe reionized in patches}

\author{Andrei Gruzinov and Wayne Hu\footnote{Alfred P. Sloan fellow}}

\affil{Institute for Advanced Study, School of Natural Sciences,
Princeton, NJ 08540}

\begin{abstract}
In a universe reionized in patches, the Doppler effect from Thomson
scattering off free electrons generates secondary cosmic microwave background
(CMB) anisotropies. For a simple model with small patches and late 
reionization, we analytically calculate the anisotropy power
spectrum. Patchy reionization can, in principle, be the main source of
anisotropies on arcminute scales. On larger angular scales, its
contribution to the CMB power spectrum is a small fraction of the primary
signal and is only barely detectable in the power spectrum
with even an ideal, i.e.~cosmic variance limited, 
experiment and an extreme model of reionization.  
Consequently patchy reionization is unlikely to affect cosmological parameter estimation 
from the acoustic peaks in the CMB.  Its detection on small angles would
help determine the ionization history of the universe in particular the
typical size of the ionized region and the duration of the reionization
process.
\end{abstract}

\section{Introduction} 
It is widely believed that the cosmic microwave background (CMB) will
become the premier laboratory for the study of the early universe and
classical cosmology.  This belief relies on the high precision of the
upcoming MAP\footnote{{\tt http://map.gsfc.nasa.gov}}
and Planck Surveyor\footnote{{\tt http://astro.estec.eas.nl/SA-general/Projects/Planck}} satellite missions and the high accuracy of theoretical
predictions of CMB anisotropies given a definite model for
structure formation (\cite{Hu95}\ 1995).  To realize the potential of the CMB,
aspects of structure formation affecting anisotropies at only the
percent level in power must be taken into account. 

A great uncertainty in models for structure formation is the extent
and nature of reionization.  
Fortunately, this uncertainty is largely not reflected in
the CMB anisotropies due to the low optical depth to Thomson
scattering at the low redshifts in question.
Reionization is known to be essentially
complete by $z \sim 5$ from the absence of the Gunn-Peterson effect in
quasar absorption spectra (\cite{Gun65}\ 1965). Significant
reionization before $z\sim 50$ will be ruled out once the tentative
detections of the CMB acoustic peaks at present are confirmed 
(\cite{Sco95} 1995). It should be possible to
deduce the reionization redshift $z_i$ through CMB polarization
measurements (\cite{Zal97}\ 1997). 

Nevertheless, the
duration of time spent in a partially ionized state will remain
uncertain. 
Moreover as emphasized by
\cite{Kai84} (1984) secondary anisotropies generated by the Doppler
effect in linear perturbation theory are suppressed on small scales
for geometric reasons (gravitational instability generates potential flows, 
leading to cancellations between positive and negative Doppler shifts).  Higher
order effects which are not generally included in the theoretical
modeling of CMB anisotropies are likely to be the main source of
secondary anisotropies from reionization below the degree scale.  Such
effects rely on modulating the Doppler effect with spatial variations
in the optical depth.  Incarnations of this general mechanism include
the Vishniac effect from linear density variations (\cite{Vis87}\
1987), the kinetic Sunyaev-Zel'dovich effect from clusters (\cite{SZ}\
1970), and the effect considered here: the spatial variation of the
ionization fraction.

Reionization commences when the first baryonic objects form stars or
quasars that convert a part of the nuclear or gravitational energy
into UV photons.  Each such source then blows out an ionization sphere around it.  Before these
regions overlap is a period when the universe is ionized in patches.
The extent of this period and the time evolution of the size and
number density of these patches depend on the nature of the ionizing
engines in the first baryonic objects. Theories of reionization do not
give robust constraints 
({\it cf.} \cite{Teg94}\ 1994; \cite{Ree96}\ 1996;
\cite{Agh96}\ 1996; \cite{Loe97}\ 1997; \cite{Hai97}\ 1997; \cite{Sil98}\ 1998;
\cite{Hai98}\ 1998). 

We therefore take a phenomenological approach to
studying the effects of patchy reionization on the CMB.  We introduce
a simple but illustrative three parameter model for the reionization
process based on the redshift of its onset $z_i$, the duration before
completion $\delta z$, and the typical comoving size of the patches
$R$. It is then straightforward to calculate the CMB anisotropies
generated by the patchiness of the ionization degree of the
intergalactic medium.

We find that only the most extreme models of reionization can produce
degree scale anisotropies that are observable in the power spectrum
 given the cosmic
variance limitations. A large signal on degree scales requires early
ionization, $z_i\gtrsim 30$, long duration, $\delta z\sim z_i$, and
ionization in very large patches, $R\gtrsim 30$Mpc. Thus the
patchiness of reionization is unlikely to affect cosmological
parameter estimation from the acoustic peaks in the CMB (\cite{Jun96}\
1996;
\cite{Zal97}\ 1997; \cite{Bon97}\ 1997).
   
On the other hand, the patchy reionization signal on the sub-arcminute
scale can, in principle, surpass both the primary and the secondary Vishniac
signals.  These may be detectable by the Planck Surveyor and upcoming 
radio interferometry measurements (\cite{Par97}\ 1997) 
if point sources can be removed
at the $\Delta T/T \sim 10^{-6}$ level. 

An explicit expression for the CMB anisotropies power spectrum
generated in a universe reionized in patches is given \S \ref{sec:explicit}. 
Simple
order of magnitude estimate of the anisotropy from patches is given in
\S \ref{sec:order}. In \S \ref{sec:power} we give a rigorous definition of our three-parameter
reionization model, and calculate the patchy part of the power
spectrum. We discuss illustrative examples in \S \ref{sec:discussion}.

\section{CMB power spectrum}
\subsection{Explicit expression}
\label{sec:explicit}
Temperature perturbations $\Delta \equiv \delta T/T$ are generated by
Doppler shifts from Thomson scattering. For small optical depths
\begin{equation}
\Delta =-\int d{\bf l}\cdot {\bf v}\sigma _Tnx_e.
\label{eqn:fundamental}
\end{equation}
All quantities here are in physical units. The integral is along the line
of site, ${\bf v}$ is the peculiar velocity of matter, $c=1$,
$\sigma_T$ is the Thomson cross section, $n$ is the number density of
free and bound electrons, $x_e$ is the local ionization fraction. 

To evaluate equation~(\ref{eqn:fundamental}) explicitly one must specify
the cosmological model.  For simplicity, we take a universe
with critical density in matter throughout; we describe the generalization
to an arbitrary FRW universe in \S \ref{sec:power}. 
We furthermore use comoving coordinates ${\bf x}$ and conformal time
$\eta\equiv \int (1+z)dt= (1+z)^{-1/2} \eta_0$, 
where $\eta_0=2/H_0$ is the present particle horizon. 
We observe at ${\bf x}=0$ and conformal time $\eta_0$
along the direction of a unit vector $\hat{\gamma}$; light propagation
is given by ${\bf x}=\hat{\gamma}(\eta _0-\eta )$. 
Equation~(\ref{eqn:fundamental})
can be
written as
\begin{equation}
\Delta (\hat{\gamma})=-\tau_0 \eta_0^3
\int {d\eta \over \eta^4} \hat{\gamma}\cdot {\bf v}[ \eta ,\hat{\gamma}(\eta _0-\eta )] x_e[ \eta ,\hat{\gamma}(\eta _0-\eta )] .
\end{equation}
Here $\tau _0\equiv \sigma _Tn_0 \eta_0$ is
the optical depth to Thomson scattering across the present particle horizon.

The
scales contributing to the peculiar velocity are still in the linear
regime, therefore
\begin{equation}
{\bf v}(\eta ,{\bf x})={\eta \over \eta _0}{\bf v}({\bf x}),
\end{equation}
where ${\bf v}({\bf x})$ is the peculiar velocity today. The final
explicit expression for the CMB temperature perturbation generated
during reionization is
\begin{equation}
\Delta (\hat{\gamma})=-\tau_0 \eta_0^2
\int {d\eta \over \eta ^3} \hat{\gamma}\cdot {\bf v}[ \hat{\gamma}(\eta _0-\eta )] x_e[ \eta ,\hat{\gamma}(\eta _0-\eta )] .
\label{eqn:temperature}
\end{equation}
The correlation function of the temperature perturbations is defined
as
\begin{equation}
C(\theta )=\left<\Delta (\hat{\gamma}_1)\Delta
(\hat{\gamma}_2)\right>|_{\hat{\gamma}_1\cdot \hat{\gamma}_2=\cos \theta}.
\end{equation}
With temperature perturbations given by equation~(\ref{eqn:temperature}), 
this becomes
\begin{eqnarray}
C(\theta )&=&\tau_0^2 \eta _0^4 \int {d\eta _1\over \eta _1^3}
\int {d\eta _2\over \eta _2^3} \big<
\hat{\gamma}_1\cdot {\bf v}({\bf x}_1)
\hat{\gamma}_2\cdot
{\bf v}({\bf x}_2)
\nonumber\\ &&\quad \times \, x_e(\eta _1,{\bf x}_1)x_e(\eta _2,{\bf x}_2)\big>,
\label{eqn:correlationgeneral}
\end{eqnarray}
where we denote ${\bf x}_1\equiv \hat{\gamma}_1(\eta _0-\eta _1)$ and
${\bf x}_2\equiv \hat{\gamma}_2(\eta _0-\eta _2)$.

\subsection{Order of magnitude estimates}
\label{sec:order}

Consider the following patchy reionization scenario. The universe was
reionized in randomly distributed patches with a characteristic
comoving size $R$. The patches appeared at random in space and
time. Once a reionized patch appears, it moves with matter. The
average ionization fraction, that is the filling fraction of fully
ionized patches, grows monotonically from $X_e=0$ at high redshifts to
$X_e=1$ at low redshifts. We consider late reionization (optical depth
to Thomson scattering is small) and small patches (smaller than the
characteristic length scale of the peculiar velocity field). We assume
that reionization occurred at redshift $z_i$, and the patchy phase
duration is given by $\delta z$.

The angular scale of the patchy signal is given by the ratio of the
size of patches to the distance to them in comoving coordinates, 
$\theta \sim R/(\eta_0-\eta_i)$. 
Assuming that the patches are uncorrelated, the spectrum of 
fluctuations should be white noise above this scale which 
agrees with the exact result as we shall see (eq.~[\ref{eqn:Cl}]). 

The rms CMB temperature fluctuation $\Delta$ on scales $\theta$
due to the patchiness can be estimated as follows. Since, by
assumption, different patches are independent, $\Delta \sim
N^{1/2}\Delta _p$. Here $N$ is the number of patches on a line of
site, $\Delta _p$ is a temperature fluctuation from one patch, $\Delta
_p\sim \tau _pv(z_i)$. Here $v(z)=(1+z)^{-1/2}v(0)$ is the rms peculiar
velocity at redshift $z$, and $\tau _p$ is the optical depth for one
patch, $\tau _p \sim (1+z_i)^2\sigma _Tn_0R$. The number of patches
$N\sim \delta \eta/R\sim (1+z_i)^{-3/2}\delta z\eta _0/R$. Collecting
all the factors, we get the following estimate for the rms
anisotropies from patches
\begin{equation}
\Delta \sim \tau_0\left<v^2\right>^{1/2}({R/\eta _0})^{1/2}(1+z_i)^{3/4} (\delta z)^{1/2},
\end{equation}
which again 
agrees with the exact result (eq.~[\ref{eqn:amplitude}], up to a dimensionless
multiplier).

\subsection{Power spectrum}
\label{sec:power}

We can factor the general expression for the temperature correlation
(\ref{eqn:correlationgeneral}) as
\begin{eqnarray}
C(\theta )&=&\tau_0^2 \eta_0^4 \int {d\eta _1\over \eta _1^3}
\int {d\eta _2\over \eta _2^3}
\left<\hat{\gamma}_1\cdot {\bf v}({\bf x}_1)\hat{\gamma}_2\cdot {\bf
v}({\bf x}_2)\right>
\nonumber \\
&&\quad\times\,
\left<x_e(\eta _1,{\bf x}_1)x_e(\eta _2,{\bf x}_2)\right>.
\end{eqnarray}
This assumes that $x_e$ and ${\bf v}$ are independent random
fields. This is not strictly correct. The ionization fraction $x_e$
must be determined by the density perturbation $\delta$, and the
density perturbation is not independent of the peculiar velocity (for
example in the linear regime $\delta =-{1\over 2} \eta \nabla \cdot
{\bf v}$). However, the ionizing radiation is presumably coming from
strongly nonlinear objects, where first stars or quasars are
lightening up. At high $z$, the length scales where the density is
nonlinear are $\ll 10$Mpc comoving, which is much smaller than the
length scales contributing to the peculiar velocity. Under the
assumption of scale separation, velocity and density (and hence $x_e$)
are indeed independent.

The correlation function for the local ionization fraction $\left<x_e x_e\right>$
is not known. Our model parameterizes the correlation function
through the the patch size $R$ and 
a mean (cosmic time - dependent) ionization
fraction $X_e(\eta )$,
\begin{eqnarray}
&&\left<x_e(\eta _1,{\bf x}_1)x_e(\eta _2,{\bf x}_2)\right> = X_e(\eta _1)X_e(\eta
_2)\nonumber\\ 
&& \qquad + [ X_e(\eta _{\rm
min})-X_e(\eta _1)X_e(\eta _2)] e^{-{({\bf x}_1-{\bf x}_2)^2\over
2R^2}}\qquad.
\label{eqn:xexemodel}
\end{eqnarray} 
Here $\eta _{\rm min}={\rm min}(\eta _1,\eta _2)$. The Gaussian
function is chosen for simplicity; it could have been any function of
the separation $x$ which equals 1 at $x=0$ and gradually turns to zero
at $x>R$. For the mean ionization fraction $X_e$ we assume a change
from 0 to 1 at a redshift $z_i$, with the transition occurring in a
redshift interval $\delta z$. We also assume $\delta z\ll
z_i$ (this is true in all of the models of reionization that we are aware of).

CMB anisotropies generated (and erased) due to the spatially constant
part of the correlation function (\ref{eqn:xexemodel}) are obviously the same as in the
model with a uniform time-dependent reionization. The spatially
varying part is responsible for generating new anisotropies; its
contribution to erasing the primary anisotropies is negligible. The
anisotropy suppression is mainly determined by the total optical depth
to Thomson scattering and is insensitive to the small-scale structure
of the ionization fraction $x_e(\eta ,{\bf x})$.

The CMB correlation function due to the patchy part only is
\begin{eqnarray}
\label{eqn:cthetaintegral}
C^{\rm (p)}(\theta ) &=& \tau_0^2 \eta_0^4\int_{0}^{\eta _0} {d\eta
_1\over \eta _1^3} \int_{0}^{\eta _0} {d\eta _2\over \eta
_2^3}I_{12} \\ && \quad\times \left<\hat{\gamma}_1\cdot {\bf
v}({\bf x}_1)\hat{\gamma}_2\cdot {\bf v}({\bf x}_2)\right>e^{-{({\bf
x}_1-{\bf x}_2)^2\over 2R^2}} \nonumber,
\end{eqnarray}
where we denote $I_{12}\equiv X_e(\eta _{\rm min})-X_e(\eta
_1)X_e(\eta _2)$.  The correlation function is non-negligible only for
$|{\bf x}_1-{\bf x}_2|\lesssim R$. By assumption, $R$ is much smaller
then the characteristic scale of the peculiar velocity field. Also
$|{\bf x}_1-{\bf x}_2|\lesssim R$ requires that the lines of sight
$\hat{\gamma}_1$ and $\hat{\gamma}_2$ be nearly parallel. Then
\begin{equation}
\left<\hat{\gamma}_1\cdot {\bf v}({\bf x}_1)\hat{\gamma}_2\cdot {\bf
v}({\bf x}_2)\right>\approx {1\over 3}\left<v^2 \right>,
\end{equation}
where $\left<v^2 \right>$ is the mean squared peculiar velocity today. For $z_i\gg
1$, the integral (\ref{eqn:cthetaintegral}) is dominated by small conformal times, and we
have $|{\bf x}_1-{\bf x}_2|^2\approx \theta ^2(\eta _0-\eta_i)^2+(\eta _1-\eta
_2)^2$. Then
\begin{eqnarray}
C^{\rm (p)}(\theta )&\approx& {1\over 3}\tau_0^2\eta_0^4\left<v^2\right>e^{-{(\eta
_0 - \eta_i)^2\theta ^2\over 2R^2}}\nonumber \\
&&\times \int_{0}^{\infty }{d\eta _1\over \eta _1^3}
\int_{0}^{\infty } {d\eta _2\over \eta _2^3}I_{12}e^{-{(\eta _2-\eta
_1)^2\over 2R^2}}.
\end{eqnarray}
We assume that $\eta _i\delta z/(1+z_i)\gg R$ (with $\eta
_i\equiv \eta (z_i)$) and that during the patchy phase,
$z_i>z>z_i-\delta z$, the ionization fraction $X_e$ grows linearly from
0 to 1 such that eventually both hydrogen and helium are fully ionized. 
Then
%
\begin{equation}
C^{\rm (p)}(\theta )=Ae^{-{\theta ^2\over 2\theta _0^2}},
\label{eqn:cthetagen}
\end{equation}
where the characteristic angular scale is
\begin{equation}
\theta _0={R\over \eta _0 - \eta_i} = {R \over \eta_0} {(1+z_i)^{1/2} \over
(1+z_i)^{1/2} - 1},
\label{eqn:angularscale}
\end{equation}
and the amplitude is
\begin{equation}
A={\sqrt{2\pi }\over 36} \tau _0^2\left<v^2\right>{R\over \eta _0}\delta
z(1+z_i)^{3/2}.
\label{eqn:amplitude}
\end{equation}
Note that a critical matter-dominated universe is assumed in this
expression.  To generalize this result replace $\eta_0-\eta_i$ by
the comoving angular diameter distance in 
equation~(\ref{eqn:angularscale}) and a factor of $(1+z_i)$ in
equation~(\ref{eqn:amplitude}) with the appropriate velocity growth
factor.

The power spectrum is given by the spherical harmonics decomposition
\begin{equation}
C_l^{\rm (p)}=2\pi \int d\cos \theta P_l(\cos \theta )w_p(\theta ).
\end{equation}
For $l\gg 1$,
\begin{equation}
C_l^{\rm (p)}\approx 2\pi \int_0^{\infty} \theta d\theta J_0(l\theta )w_p(\theta
)=2\pi A\theta _0^2e^{-{\theta _0^2l^2\over 2}}.
\end{equation}
The power per octave is
\begin{equation}
{l^2C_l^{\rm (p)}\over 2\pi }=Al^2\theta _0^2e^{-{\theta _0^2l^2\over 2}}.
\label{eqn:Cl}
\end{equation}

The anisotropy power reaches the maximal value
\begin{equation}
({l^2C_l^{\rm (p)}\over 2\pi })_{\rm max}={\sqrt{2\pi }\over 18{\rm e}} \tau
_0^2\left<v^2\right>{R\over \eta _0}\delta z(1+z_i)^{3/2},
\end{equation}
at
\begin{equation}
l_{\rm max}={\sqrt{2} \eta _0\over R} [1 - (1+z_i)^{-1/2}].
\end{equation}

\subsection{Discussion}
\label{sec:discussion}

The signal from patchy reionization in our model depends on four
quantities: the rms peculiar velocity 
$\left< v^2 \right>^{1/2}$ today, the redshift of reionization $z_i$,
its duration $\delta z$ and the characteristic comoving size of the patches $R$.
The structure formation model specifies
the power spectrum of fluctuations which in turn tells us the 
rms peculiar velocity. Let us now consider the patchy reionized signal in 
the context of a specific model for structure formation.  

For illustrative purposes, let us consider
a cold dark matter model with $h=0.5$, $\Omega_b  =0.1$, and a scale-invariant $n=1$ 
spectrum of initial fluctuations.  Normalizing the spectrum to the
COBE detection via the fitting formulae of \cite{Bun97} (1997) (their
equations [17]-[20]) and employing the analytic fit to the transfer function 
of \cite{Eis98} (1998) (their equations [15]-[24])
we find an rms velocity  
of $\left<v^2\right>^{1/2}=3.9\times 10^{-3}$.  
With the present optical depth of $\tau_0 = 0.122 \Omega_b h = 0.0061$,
we have a maximal anisotropy of
\begin{equation}
({l^2C_l\over 2\pi })_{\rm max}=2.41\times 10^{-15}{R\over {\rm
Mpc}}\delta z(1+z_i)^{3/2},
\end{equation}
at
\begin{equation}
l_{\rm max}={16958\over R/{\rm Mpc}} [ 1 - (1+z_i)^{-1/2}].
\end{equation}

The power spectrum of the model in principle also tells us the
remaining parameters of the ionization: its redshift $z_i$,
duration $\delta z$ and typical patch size $R$.
Unfortunately, these quantities depend on details of the cooling and
fragmenting of the first baryonic objects to form the ionizing 
engines.  We therefore consider $5 \lesssim z_i \lesssim 50$ which spans 
the range of estimates in the literature
(\cite{Teg94}\ 1994; \cite{Ree96}\ 1996;
\cite{Agh96}\ 1996; \cite{Loe97}\ 1997; \cite{Hai97}\ 1997; \cite{Sil98}\ 1998;
\cite{Hai98}\ 1998).  Reionization, once it commences, is generally completed in
a time short compared with the expansion time at that epoch $\delta z /(1 +z_i) 
< 1$ by the coalescence of patches that are small compared with the 
horizon at the time $R/\eta_i
\ll 1$ at the time.  Again the exact relations depend on the efficiency with
which the first objects form and create ionizing radiation 
(see e.g. \cite{Teg94}\ 1994). 

\begin{figure}[htb]
\psfig{figure=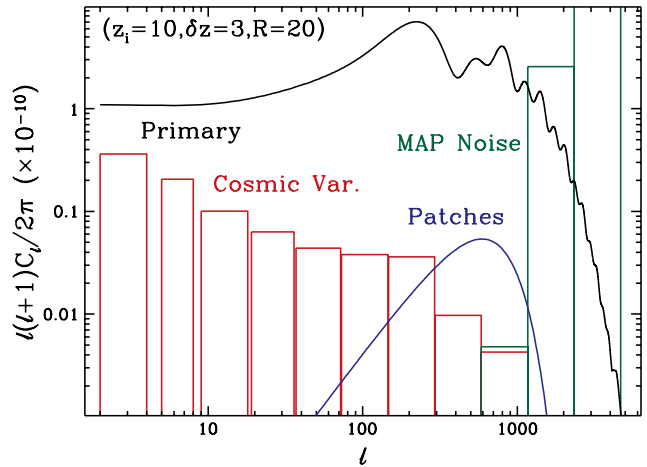,width=3.3in}
\caption{CMB anisotropy power spectra in a CDM model with extreme patchiness. 
Shown here are the primary anisotropy 
and the patchy reionization anisotropy, eq.~(\protect\ref{eqn:Cl}) with
$z_i=10$, $\delta z=3$, $R=20$Mpc. These signals are compared with
the cosmic variance of the primary anisotropy and the noise of
the MAP satellite (in logarithmic bins).}
\label{fig:patch10}
\end{figure}

Let us consider an extreme example of $z_i=10$, $\delta z=3$, $R=20{\rm Mpc}$. 
Then the maximal
power is $\approx 5.3\times 10^{-12}$ at $l\approx 590$, the primary
signal at these scales is $\approx 3\times 10^{-10}$ -- the
contribution of the patchy reionization is small in comparison 
(see Fig.~\ref{fig:patch10}).  However, in light of 
the high precision measurements expected from the MAP and Planck satellites 
such a signal is not necessarily negligible.  
The ultimate limit of detectability through
power spectrum measurements is provided by so-called cosmic variance.
This arises since we can only measure $2\ell+1$ realizations
of any given multipole such that power spectrum estimates will vary by
\begin{equation}
\delta C_\ell = \sqrt{2 \over 2\ell+1} C_\ell^{(\rm primary)}.
\end{equation}
Detection of a broad feature such as that from patchy reionization is assisted
in that we may reduce the cosmic variance by averaging over many $\ell$'s.  
We show an example of this averaging in Fig.~\ref{fig:patch10} (lower left boxes).
In this model, the patchy reionization signal can be detected at the several
$\sigma$ level if cosmic variance were the main source of uncertainty.  
Of course a realistic experiment also has noise and systematic errors.
We also show the noise error contributions expected from the MAP experiment
in Fig.~\ref{fig:patch10}.     

An important additional source of uncertainty is provided by other
unknown aspects of the model.  Indeed it is hoped that the CMB power spectrum
can be used to measure fundamental cosmological parameters to high precision.
Excess variance from patchy reionization can in principle cause problems for
cosmological parameter estimation from the CMB if not included in the model.
It would remain undetected and produce parameter misestimates if
its signal can be accurately mimicked by variations in the other parameters.
Fortunately, the angular signature we find here -- $\ell^2$ white noise until some
cut off due to the patch size -- does not resemble the signature of other cosmological
parameters which alter the positions and amplitudes of the acoustic peaks 
(see \cite{Bon97}\ 1997; \cite{Zal97}\ 1997).  Coupled with the small amplitude of the effect on the 10 arcminute to degree scale for even this extreme model, it is unlikely that patchy reionization will significantly affect parameter
estimation through the CMB. 

We have called the ($z_i=10,\delta z=3,R=20$) model extreme, because of the size of patches; the reionization redshift and duration would be considered reasonable by a number of theories. For example the early quasar model of Haiman \& Loeb (1998) does predict $z_i\sim 10$ and $\delta z \sim 3$. However, their ``medium quasar'' emits only $\sim 10^{67}$ ionizing photons during its life time. These photons cannot ionize a bubble larger than $R\sim 1$Mpc comoving.

\begin{figure}[htb]
\psfig{figure=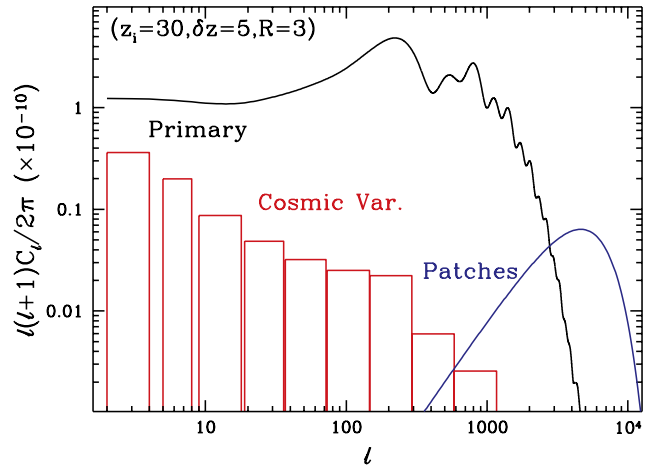,width=3.3in}
\caption{CMB anisotropy power spectra in a CDM model with early reionization. 
Shown here are the primary anisotropy suppressed by rescattering 
and the patchy reionization anisotropy, eq.~(\protect\ref{eqn:Cl}) with
$z_i=30$, $\delta z=5$, $R=3$Mpc. These signals are compared with
the cosmic variance of the primary anisotropy achievable by an ideal
experiment in the absence of galactic and extragalactic foregrounds.}
\label{fig:patch30}
\end{figure}

Perhaps more interesting is the case where reionization takes 
place at a higher redshift with
for example 
$z_i=30$, $\delta z=5$, $R=3{\rm Mpc}$.  The reduction in the patch size
causes the signature to move to smaller angles where the primary signal
is negligible due to dissipational effects at recombination.  
The increase in the optical depth at this higher redshift is counterbalanced
by the reduction in the rms fluctuation due to the number of patches along
the line of site such that the amplitude of the signal increases
only moderately.  Here the maximal 
power is $\approx 6.2\times10^{-12}$ at $l\approx 4650$ 
(see Fig.~\ref{fig:patch30}).  Patchy reionization effects exceed 
the Vishniac signal at these
scales ($\approx 3\times 10^{-12}$) which is believed to be
the leading other source of secondary anisotropies 
(\cite{Hu96}\ 1996).  

Although the morphology and amplitude 
of the patchy reionization and Vishniac signals are similar, the Vishniac
effect is fully specified by the ionization redshift and the spectrum
of initial fluctuations and hence may be removed once these are determined
from parameter estimation at larger angular scales.  Likewise, 
since the rms peculiar velocity $\left< v \right>$ and the ionization
redshift $z_i$ will be specified by the large scale observations, 
the amplitude of the signal can be used to estimate
the duration of reionization $\delta z$ and its angular location 
the typical comoving size of the bubbles $R$. 

In summary, the patchiness of reionization leaves a potentially observable 
imprint on the CMB power spectrum, but one that is unlikely to affect cosmological
parameter estimation from the acoustic peaks in the CMB.   We show how the signature
scales with the gross properties of reionization -- its redshift, duration, and
typical patch size.   Observational detection of this signature would provide 
useful constraints on the presently highly uncertain reionization scenarios but
will likely require experiments with angular resolution of an 
arcminute or better and foreground subtraction at better than the 
$\delta T/T \sim 10^{-6}$ level.

\acknowledgements
We thank 
R. Juszkiewicz, A. Liddle, M. Tegmark and M. White
for discussions.  This work
was supported by NSF PHY-9513835.  WH was also supported by the
W. M. Keck Foundation.

\end{document}